# Sensitivity of ferromagnetic resonance in PdCo alloyed films to hydrogen gas

C. Lueng[1], P. Lupo[2], P.J. Metaxas[1], A.O. Adeyeye[2] and M. Kostylev[1]

*1. School of Physics and Astrophysics M013, The University of Western Australia, 6009 Crawley, WA, Australia*

*2. Department of Electrical and Computer Engineering, National University of Singapore, Singapore*

Abstract: In this work we studied the ferromagnetic resonance response of $Co_xPd_{1-x}$ alloy samples with different cobalt contents (x = 0.65, 0.39, 0.24 & 0.14). We found significant differences in the response of the samples to the presence of hydrogen gas in the samples' environment. Two particular films (with x=0.39 and 0.24) demonstrated behaviour which is promising for application in hydrogen gas sensing. Using the $Co_{39}Pd_{61}$ alloy thin film, we were able to measure hydrogen gas concentration in a very broad range (from <0.1% to 100%) at a fixed value of the external magnetic field. Finally, we demonstrate that the $Co_{24}Pd_{76}$ alloy thin film is sensitive to ultra-low hydrogen gas concentrations – from 10 to $10^4$ ppm.

1. Introduction

Recently, research on hydrogen gas sensors [1] has become critical due to the potential to use hydrogen gas as a sustainable and renewable energy source. Hydrogen gas ($H_2$) is flammable and explosive. However, it cannot be detected by human senses. As such, reliable, fire-safe and high-sensitivity hydrogen gas sensors (HGS) are needed to monitor $H_2$ concentration in the environment. The requirements for the range of detection depends on the application [2-4]. Safety sensors should cover the range from 0.01% to 10% of $H_2$ in the environment while the detection range for monitoring fuel cell performance should span from 1% to 100%. In our previous works we reported a magnetic hydrogen gas sensor (mHGS) approach [5-7]. The concept is based on the use ferromagnetic resonance (FMR) for detecting small variations of magnetic anisotropy of ferromagnetic materials. Specifically, with FMR we measure a change in the strength of the perpendicular magnetic anisotropy (PMA) existing at the interface between palladium (Pd) and cobalt (Co) layers in bi-layer Pd/Co thin films. The change is induced when palladium absorbs hydrogen gas [8-10].

We found that this approach can be used to cover a very large range of $H_2$ concentrations – from 0.2% to 100%. To enable this large sensitivity range, the range was divided into four sub-ranges and the magnetic field applied to our device-prototype adjusted to maximise the device sensitivity to one of those sub-ranges [11]. We speculated that the whole range could be covered by employing three or four sensors operating in parallel, each tuned to a specific concentration sub-range.

Two characteristics of the FMR response of the films are important for the mHGS functionality. One is the magnitude of the FMR peak (field or frequency) shift induced by $H_2$. The second important parameter is the absolute value of the FMR peak height ("FMR absorption amplitude" in our measurement method). A large amplitude response will enable simpler electronics for FMR absorption detection.

In order to maximise the $H_2$ induced frequency shift, the PMA should be maximised. This is because the stronger the PMA, the larger absolute variation of its strength one can expect in the presence of the same amount of $H_2$ in the environment. In FMR measurements, the *interface* PMA manifests as an effective bulk-like field whose strength scales as $1/L$, where $L$ is the thickness of the ferromagnetic layer [12]. Accordingly, one may expect a reduction of the $H_2$-induced FMR peak shift with an increase in $L$. Indeed, our recent experiment confirmed that the sensitivity of a bilayer film based mHGS to hydrogen gas is inversely proportional to the thickness of the film's cobalt layer [13].

On the other hand, the FMR absorption amplitude scales as the volume of the resonating material and, consequently, as $L$. Accordingly, a decrease in $L$ leads to a decrease in the signal-to-noise ratio at the HGS output. Previously, work using benchtop detection electronics, we found that $L$=5nm was an optimum thickness. On one hand, it enables a significant FMR field shift in the presence of $H_2$; on the other hand, the strength of the FMR absorption is still large enough in order to detect it with a lock-in based FMR setup in a laboratory setting [14]. Note that for this value of $L$ PMA in Pd/Co bilayers is weaker than the shape anisotropy and the static magnetization vector for the film lies naturally in the film plane.

A further improvement in the amplitude of the FMR absorption is needed however to make our HGS concept relevant for real-world application. In Ref. [6] we proposed how this can be achieved by optimising the broadband FMR configuration used to probe the FMR response of bilayer materials. Here we report on an alternative approach which is based on a $H_2$-sensitive alloy.

Recently it has been demonstrated that the strength of PMA in CoPd *alloy* films is sensitive to the presence of $H_2$ [15-17]. Since a bulk PMA is present in these systems (rather than an interfacial PMA), one may expect that its effect on the FMR field or frequency will not drop with an increase in the thickness of the film. This could remove the limitation for the maximum thickness of the ferromagnetic layer, thus allowing an increase in the FMR absorption amplitude without sacrificing the material's sensitivity to $H_2$ (i.e. the magnitude of the $H_2$-induced FMR peak shift). In this work, we put this idea to test by investigating the FMR response of CoPd alloy films in the presence of $H_2$.

2. Material and Methods

15 nm thick, continuous $Co_xPd_{1-x}$ alloys films with different Co concentrations $x$ were prepared using co-sputtering. Four samples were fabricated with x = 0.65, 0.39, 0.24 & 0.14. To carry out the FMR characterisation of the films in the presence of hydrogen gas, a pressure-tight chamber with a 0.3mm-wide microstrip line on its floor is used [5,7,13]. During the measurements, the film is facing the microstrip line (i.e. the film's substrate is facing away from the stripline) in order to maximize the absorption signal. A static magnetic field is applied to the film in order to yield a finite resonance frequency. The field is applied perpendicular to the film plane (PP configuration). This is done because we previously demonstrated that the $H_2$–induced FMR field shift is maximised in the PP configuration [7]. A field-modulated FMR method is employed to improve the signal to noise ratio [14]. By applying this method, instead of measuring an absorption peak which would normally have a Lorentzian shape, the first derivative of a Lorentzian is measured. A microwave interferometric receiver is also employed to further improve the signal to noise ratio [18]. More details on the broadband FMR method can be found in the recent review [14].

To prepare a broad range of hydrogen gas concentrations, we employ a custom-built gas mixing setup which utilises Alicat® mass flow controllers which are computer-controlled using the Flow Vision ® MX software. The prepared gas mixture flows through the pressure-tight chamber. All FMR measurements are carried out under a constant gas flow rate of 800 sccm and atmospheric pressure. Because a very wide range of hydrogen gas concentrations is probed (0.1% to 100%), nitrogen is used as carrier gas for the mix. This ensures the fire safety of our experiment - an air-H$_2$ gas mix is flammable above 4% of H$_2$ concentration. Note that in Ref. [6] we looked at the carrier gas effect and found that the responses in air and in nitrogen as carrier gases converge when approaching the 2% concentration point from below.

3. Experimental results

The PP FMR configuration is also convenient for extraction of magnetic parameters for the samples. This is because, following the Kittel Equation for the PP configuration

$$\omega = \gamma(H - 4\pi M_{eff}) \quad (1)$$

the FMR frequency $\omega$ scales linearly with the effective magnetisation $4\pi M_{eff} = 4\pi M_s - H_{PMA}$ which is the difference between the saturation magnetisation $4\pi M_s$ and the effective field of PMA. The slope of the $\omega(H)$ dependence is given by the gyromagnetic ratio for the material under test $\gamma$. The intercept of the straight line with the vertical axis gives the value of $4\pi M_{eff}$ for it. From Fig. 1, one sees that the lower the Co concentration in the alloy, the lower 4piM$_{eff}$. Also, the gyromagnetic ratio grows with the decrease in *x*.

The FMR data also reveal that the static magnetisation vector lies in the film plane for the largest x (Co$_{0.65}$Pd$_{0.35}$), but for the other three samples the magnetic ground state the static magnetisation vector may be canted with respect to the film plane.

The evidence for this is contained in the in-plane FMR data. Whereas Sample *x*=0.65 shows a pronounced FMR response when the static field is applied in the sample plane, we could not find in-plane FMR responses for the other samples within the applied-field range available with our electromagnet (0 to 13 kOe).

Another important observation is that the resonance fields for the alloys are smaller than for pure cobalt films and decrease with a decrease in Co content [19]. As seen from Table 1, this translates into smaller effective saturation magnetization values for the alloys with smaller cobalt concentrations. This decrease in the FMR field is beneficial for applications of these materials in the mHGS sensors. Application of a large magnetic field $H = 4\pi M_s - H_{PMA}$ is required in order to magnetically saturate a film perpendicular to its plane and thus to create conditions for existence of a PP FMR response (see Eq.(1)). The decrease in the effective magnetization with the decrease in Co content translates in smaller applied fields. From Fig. 1 one sees that the FMR fields for Samples x=0.39, 024, and 0.14 are already low enough that compact permanent magnets could be used to generate the required magnetic field.

We now examine the response of the samples to hydrogen gas. Firstly, it must be noted that we found that the exposure of the Co$_{0.65}$Pd$_{0.35}$ alloy sample to H$_2$ led to irreversible changes to its FMR response. After exposure of the sample to 50% H$_2$ and then to pure N$_2$ the FMR peak did not return to the position observed for the virgin sample. Furthermore, we could not find an FMR response for x=0.14 in the presence of hydrogen gas. Therefore these two samples were excluded from further experiments and below we focus on the two remaining samples - with *x*= 0.39 (Sample 2) and 0.24 (Sample 3). Note that the resonance linewidths for the two samples as measured in their virgin states are quite different. The linewidth for *x*=0.24 is 252 Oe, whereas it is almost 3 times bigger for *x*=0.39 (692 Oe).

The samples *x*= 0.39 and 0.24 were characterized by repeatable FMR responses when cycled between pure nitrogen and hydrogen containing atmospheres. Panels (a) and (b) of Fig. 2 display families of raw FMR traces for the samples taken for different hydrogen gas concentrations. The magnitudes of the peak shifts extracted from the raw traces are shown in Fig. 3(a). Fig. 3(b) displays the shift magnitudes relative to the respective resonance linewidths.

These data demonstrate that the absolute values of the hydrogen-induced peak shift are much bigger for *x*=0.24 than for *x*=0.39 for the same hydrogen gas concentrations. The shift exceeds 10 times the linewidth for the highest $H_2$ concentration and remains larger than the linewidth at the lowest concentration of 0.1%. For *x*=0.39, the peak shift is always smaller than the linewidth.

From Fig. 2b one sees that the response amplitude for *x*=0.24 can exceed 30 microvolts. This magnitude is 6 times larger than that observed in the bi-layer Pd/Co film with Co layer thickness of 5 nm [11] (5μV). However, the concentration of cobalt in this alloy is just 24% of that in the bi-layer film. Hence, accounting for the 3 times larger thickness (15nm) of the alloyed film, the FMR response per Co atom is

$$\frac{1}{(3 \cdot 0.24)} \frac{30 \mu V}{5 \mu V} = 8.3$$

times larger for the alloyed sample than for the bi-layer film (if we neglect potential differences in crystal lattices for the two materials).

The absorption amplitude for *x*=0.39 is about 6 μV. Furthermore, the Co content in this alloy is 1.5 times larger than for Sample 3. Therefore, relative to the reference Pd/Co film, the FMR absorption amplitude per one Co atom for the alloyed film is

$$\frac{1}{(3 \cdot 0.39)} \frac{6 \mu V}{5 \mu V} = 1.1.$$

This is 8 times smaller than the *x*=0.24 film. However, this sample is characterised by a resonance linewidth, $\Delta H$, which is about three times larger. For the same amplitude of the microwave driving field, the absorption amplitude scales as $1/\Delta H$. Therefore, if $\Delta H$ for this sample were the same as for x=0.24 the absorption amplitude for it would be 3 times larger than for the reference Pd/Co bi-layer film. Hence, reducing the FMR linewidth for this alloy to the level of *x*=0.24 will be very beneficial.

On the other hand, having a material with a large $\Delta H$ and relatively strong FMR response (6 μV, as for *x*=0.39) can also be useful for magnetic $H_2$ sensing. Before we proceed to a demonstration of this, we first discuss the response time under exposure of the sample to hydrogen gas ("response time") and the time of the FMR signal recovery to its original state upon evacuation of $H_2$ evacuation from the chamber ("recovery time"). These two parameters are very important for hydrogen-gas sensitive materials [2]. We will discuss this aspect using data obtained for x=0.24. In order to measure the response and recovery times, a "time-resolved" FMR method [11] is employed. Figure 4 shows a time-resolved FMR trace taken at 10GHz with a bias magnetic field of 4160 Oe. In order to obtain this trace, the FMR frequency and the bias field were kept constant throughout the experiment and the FMR absorption amplitude was measured as a function of time. The dashed line in the plot shows hydrogen concentration as a function of time. Originally the sample environment was pure N$_2$; at 480 seconds into the experiment the atmosphere was replaced with one containing 0.1% H$_2$. At 1800 seconds the chamber was flushed with pure nitrogen gas which was then kept flowing until the end of the measurement.

As seen from Figure 1(b) the FMR peak shift for this material in the presence of just 0.1% H$_2$ is huge. Accordingly, the change in the amplitude of the differential absorption signal, once H$_2$ has been let into the chamber is very large – 30 microvolts (Figure 4). One more important observation from this figure is that although there is a sharp response of the film to the presence of hydrogen gas (the slope of the ascending section of the curve is very steep), the recovery of the system is slow (the descending section is much flatter), once hydrogen has been removed from the chamber. The response and recovery times are on the order of hundreds and thousands seconds respectively for this small concentration of H$_2$ (nearly 280min are needed to recover to 90% of the origin signal level). These large times are typical for continuous films [6,15], especially for small hydrogen gas concentrations. With an increase in the concentration, the both characteristic times drop significantly [7] (which is also seen from Fig.5). One more observation from the time-resolved measurements is that the response and recovery for the *x*=0.39 sample is much faster (not shown) than for x=0.24.

Let us now turn to the role of the resonance linewidth in the functionality of a film as an efficient mHGS. In Ref.[11] we showed that it was possible to cover a wide range of hydrogen gas concentrations with a single sensor device-prototype by tuning sensor sensitivity to different sub-ranges of the range. The sensor tuning proceeds by adjusting a static magnetic field applied to the film. This tuning was required because the hydrogen induced FMR field shift exceeded the resonance linewidth. This implies that, if a material is found which demonstrates the same absorption amplitude and the same H$_2$-induced FMR peak shift but possesses a $\Delta H$ value which is equal or larger than the maximum shift, no field adjustment will be needed in order to cover a broad H$_2$ concentrations range. This is actually the case of *x*=0.39. Figure 5 shows a time-resolved trace for this sample taken while cycling the atmosphere in the chamber between pure N$_2$ and different concentrations of H$_2$. One sees that the material responds to a change in the gas environment within minutes. Figure 5(a) displays the time-resolved trace for the concentration range from 1% to 100%, while Fig. 5(b) shows one for smaller concentrations - from 5% to 0.1%. The maxima of the negative peaks in Figs. 5(a) and (b) correspond to H$_2$-containing environments and the baseline is for pure nitrogen. Figure 5(c) displays the peak height with respect to the baseline. From this figure one sees that hydrogen concentrations as low as 0.1% can be detected with Sample 2. Furthermore, the sensitivity to hydrogen does not before reaching a H$_2$ gas concentration of 100%.

In Fig. 6 we compare the sensitivity, *S*, to H$_2$ for different materials extracted from the data in Fig. 5. We define *S* as variation in the absorption amplitude per unit of concentration (measured in µV/%). Two samples are considered – x=0.39 and the Pd/Co bilayer film studied in [11]. From this figure one sees that Sample 2 (*x*=0.39) may cover the whole range accessible with the bi-layer film with the level of sensitivity across the range which is the same as for the bi-layer film. However, the applied field had to be re-adjusted 2 times in order to cover this concentration range with the bi-layer film. Within contrast, for the alloyed film no field re-adjustment is needed. This is because the alloyed sample is characterised by a combination of a strong FMR absorption, a relatively large H$_2$-induced peak shift, and a large FMR linewidth.

Figure 7 displays the same sensitivities but plotted on a logarithmic scale. One sees that the *S* vs. concentration traces are perfectly linear on that scale. The solid lines in the figure are respective linear fits. One sees that the intercepts of these lines with the horizontal axis is located at very low concentrations, especially for *x*=0.24. This suggests that the alloyed samples may be prospective candidates for detection of low hydrogen gas concentrations as well. However, this is out of scope of our paper, as reaching such low concentrations is not possible with the current version of our gas preparation equipment.

4. Conclusions

In this study, we explored the potential of CoPd alloyed films as an active medium for the Ferromagnetic Resonance based hydrogen gas sensing. We found that this system allows one to measure hydrogen gas concentration in a very broad concentration range, including the vicinity of the 100% concentration mark. Detection is based on modification of the strength of the effective field of the bulk perpendicular anisotropy, unlike the previous FMR-based studies where interface PMA was employed. We obtained the same amplitude of the FMR response for the alloyed materials as for the bi-layer films. However, the saturation magnetisation for the alloys is significantly lower than effective magnetisation of the 5nm-thick pure-cobalt layers used in [7] which leads to significantly smaller magnetic fields being required in order to magnetically saturate the samples perpendicular to their plane. Such fields can be applied with inexpensive permanent magnets. This is very advantageous, as it has been previously shown [7] that the perpendicular to plane magnetisation leads to a significant increase in sensitivity of the FMR response of the magnetic films to hydrogen gas.

Acknowledgements

Funding by the Australian Research Council, the University of Western Australia within the frameworks of the "Near Miss", "Research Collaboration Award", and "Teaching Relief" and from the National Research Foundation, Prime Minister's Office, Singapore under its competitive Research Programme (CRP Award No. NRF-CRP 10-2012-03) is acknowledged. P.J.M is the recipient of an Australian Research Council Discovery Early Career Researcher Award (project number DE120100155).8/*

Table 1. Magnetic parameters for the films extracted from the data in Fig. 1.

| Co content | Pd content | $\gamma$ (MHz/Oe) | $4\pi M_{eff}$ (Oe) |
|---|---|---|---|
| 0.65 | 0.35 | 0.00284 | 9518.6 |
| 0.39 | 0.61 | 0.00299 | 2050.4 |
| 0.24 | 0.76 | 0.00315 | 922.3 |
| 0.14 | 0.86 | 0.00610 | 36.6 |

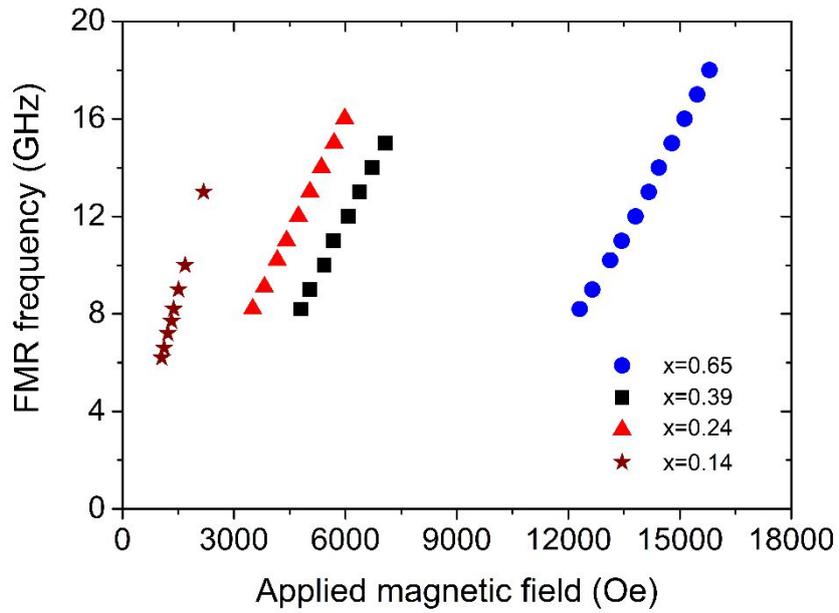

Fig. 1. FMR frequency as a function of applied magnetic field.

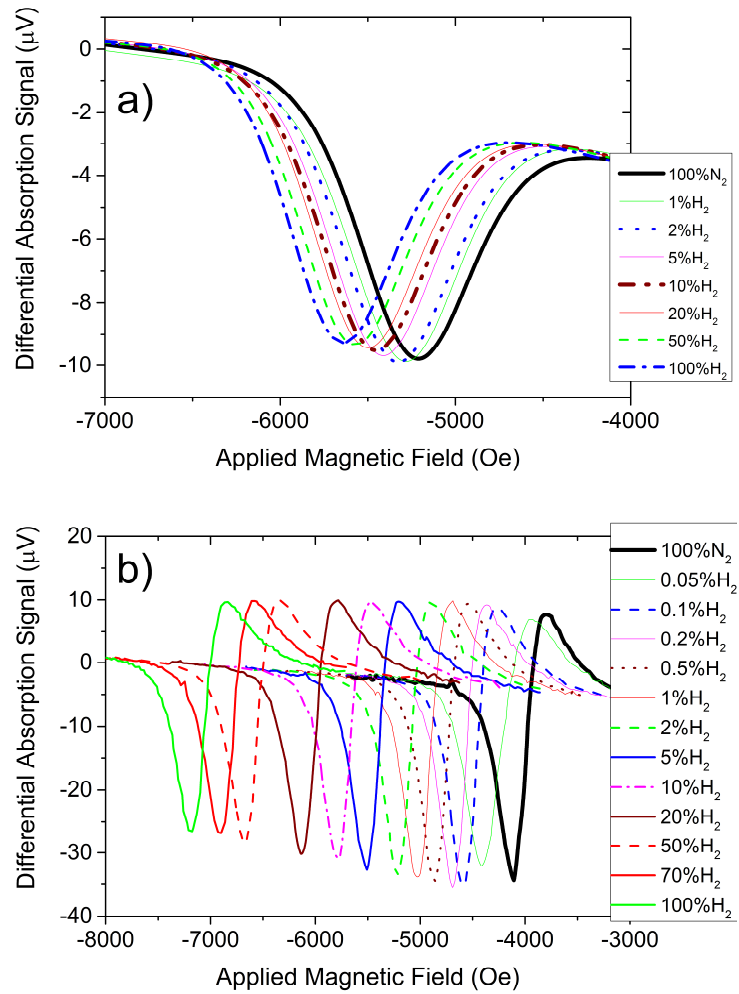

Fig. 2. FMR Traces for different concentrations of hydrogen gas for (a) *x*=0.39 and *x*=0.24 .

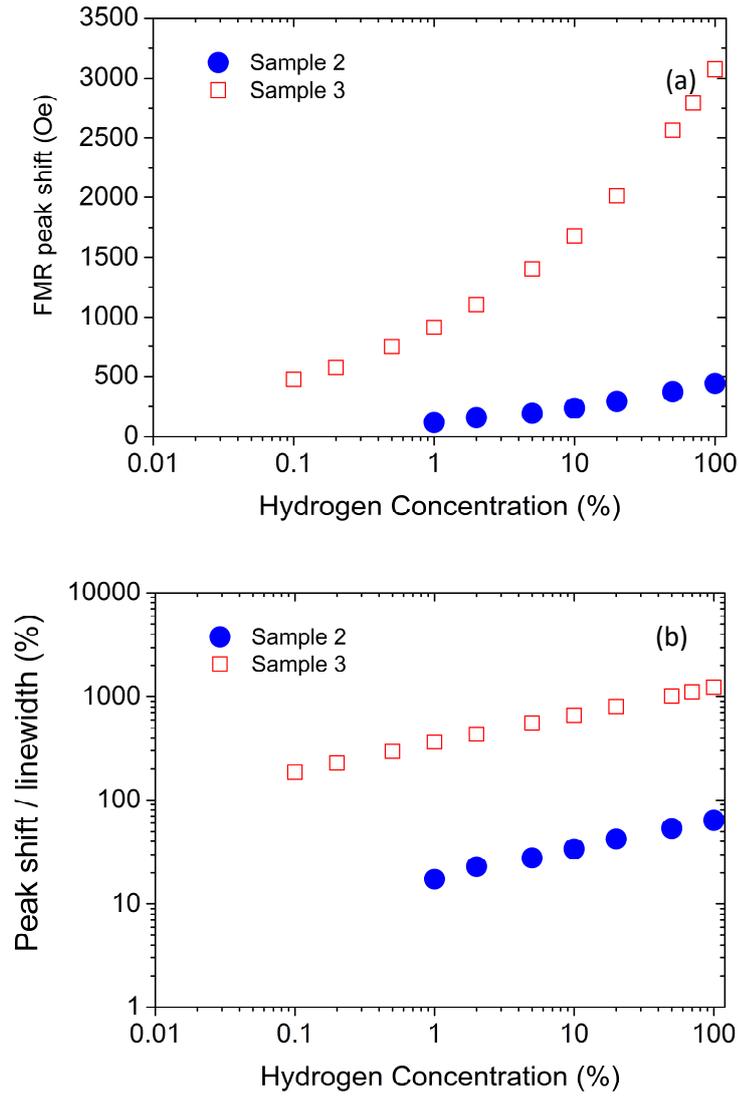

Fig. 3. (a) Hydrogen-induced FMR peak shifts for the two samples from Fig. 2. (b) Ratio of the FMR peaks shift to the respective linewidths (b).

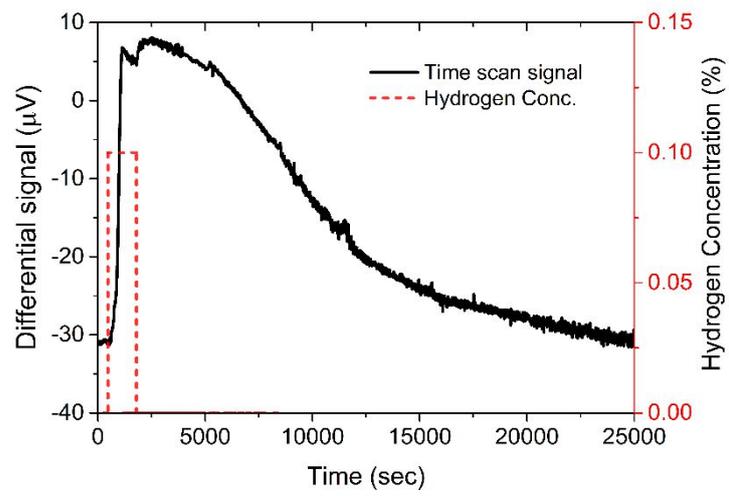

Fig. 4. Responses and recovery time for x=0.39 and 0.1 of $H_2$ in the gas mix.

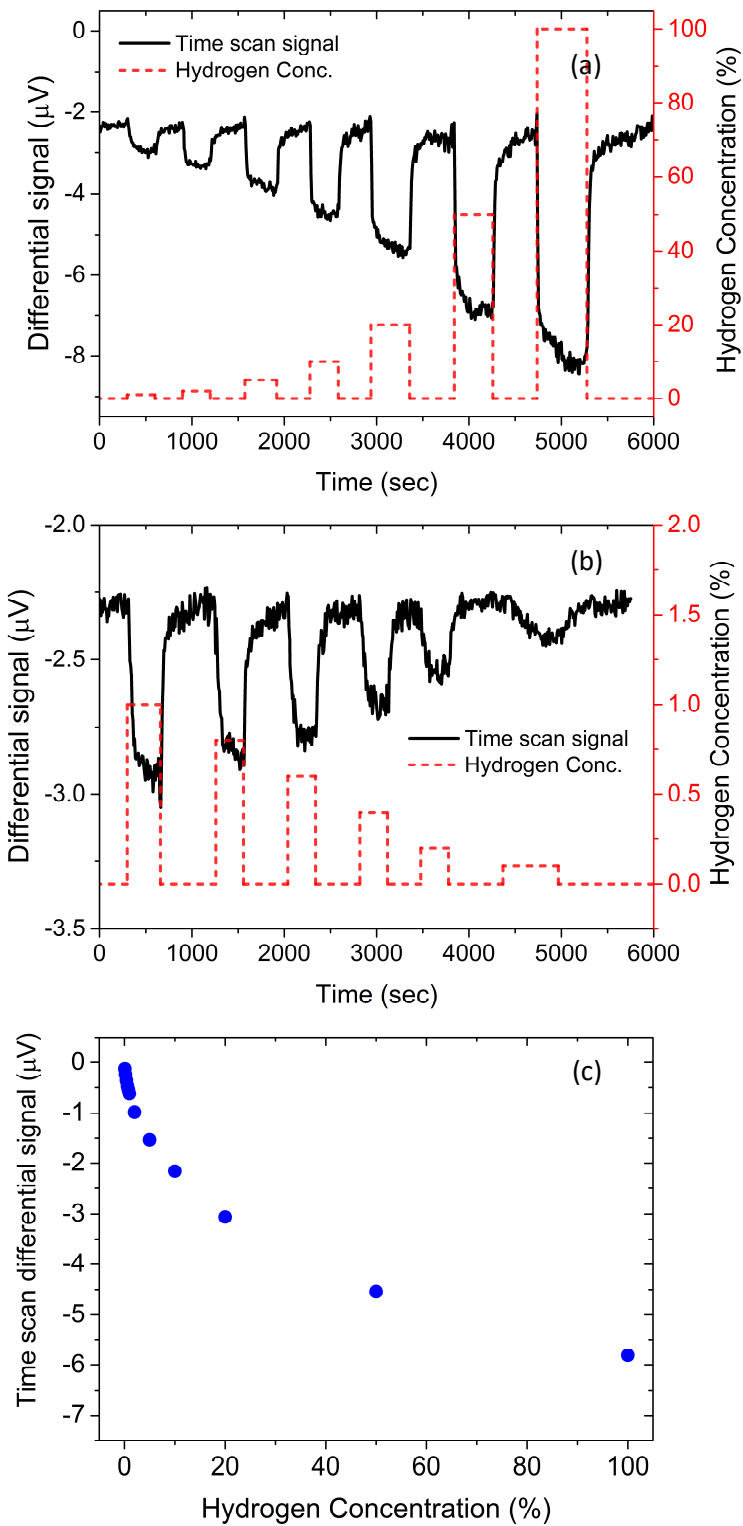

Figure 5. (a) Black solid line: time-resolved FMR for *x*=0.39 trace taken while cycling between pure nitrogen gas (positive peaks) and different concentrations of hydrogen gas in nitrogen as a carrier gas (negative peaks). Hydrogen gas concentrations are from 1% to 100%. Red dashed line: respective concentration of hydrogen (right-hand vertical axis). (b) As for (a) but for 1% to 0.1% hydrogen gas. Applied field is 5809 Oe and the frequency is 10 GHz. (c) Variation of the FMR absorption amplitude as a function of hydrogen gas concentration. These points were extracted from the data in Panels (a) and (b) by measuring the heights of the negative peaks.

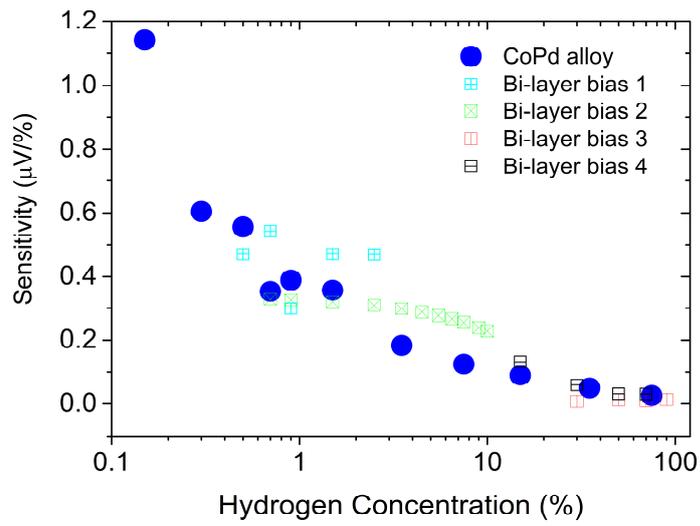

Fig. 6 Sensitivity to hydrogen gas for the CoPd alloyed film (x=0.39, large blue circles) and a Pd/Co bi-layer film from Ref [11] (squares).

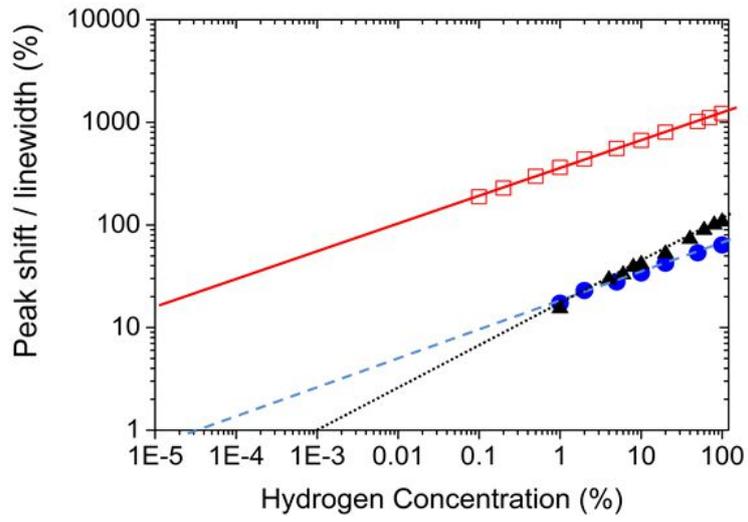

Fig. 7 Film sensitivity to hydrogen gas plotted on a logarithmic scale. Dots: experiment; lines: linear fit to the experimental data. Squares and solid line: x=0.24. Circles and dashed line: *x*=0.39. Triangles and dotted line: bi-layer sample from Ref.[11].